\documentclass[12pt]{article}
\usepackage{hyperref}
\usepackage{amssymb,amsmath,graphicx,latexsym}

\begin{document}
\title{A Logarithmic Correction in the Entropy Functional Formalism}
\author{Fay\c{c}al Hammad$^{1}$\thanks{fhammad@ubishops.ca} and Mir Faizal$^{2}$\\
$^{1}$\emph{\small{Physics Department \& STAR Research Cluster, Bishop's University, and}}\\
\emph{\small{Physics Department, Champlain College-Lennoxville}}\\
\emph{\small{2600 College Street, Sherbrooke, Qu\'{e}bec J1M 1Z7, Canada}}\\
$^{2}$\emph{\small{Department of Physics and Astronomy, University of Waterloo,}}\\
\emph{\small{Waterloo, Ontario N2L 3G1, Canada}}}


\date{}
\maketitle
\begin{abstract}
The entropy functional formalism allows one to recover general relativity, modified gravity theories, as well as the Bekenstein-Hawking entropy formula. In most approaches to quantum gravity, the Bekenstein-Hawking's entropy formula acquires a logarithmic correction term. As such terms occur almost universally in most approaches to quantum gravity, we analyze the effect of such terms on the entropy functional formalism. We demonstrate that the leading correction to the micro-canonical entropy in the entropy functional formalism can be used to recover modified theories of gravity already obtained with an uncorrected micro-canonical entropy. Furthermore, since the entropy functional formalism reproduces modified gravity, the rise of gravity-dependent logarithmic corrections turns out to be one way to impose constraints on these theories of modified gravity. The constraints found here for the simple case of an $\mathcal{F}(R)$-gravity are the same as those obtained in the literature from cosmological considerations.
\end{abstract}

\begin{quote}
\end{quote}



\section{Introduction}\label{sec:1}
The origin of black hole thermodynamics was originally discovered using using semi-classical involving quantum field theory on curved spacetime \cite{Bekenstein, Hawking}. In this analysis the gravitational field was described by classical general relativity. Therefore, any new revealed aspect of black hole thermodynamics would certainly shed more light on the long-sought quantum theory of gravity.

One of the main results in black hole thermodynamics, on which many of the approaches to quantum gravity seem to agree \cite{Strominger, Ashtekar, Carlip, Solodukhin, Kaul, Das1}, is the form of the leading correction brought to the famous Bekenstein-Hawking entropy formula. Bekenstein and Hawking originally found that the entropy of any large black hole follows the area-law, that is, entropy is proportional to the area of the event horizon. What recent research showed is that this area-law is usually only an approximation because, when either stringy effects or loop quantum gravity's geometry are taken into account, a logarithmic term arises in addition to the usual linear area-dependence found by Bekenstein and Hawking. But then, the question that naturally rises is why do different approaches recover the same form for the corrections, and what is the origin of this 'universal' logarithmic correction.

Relying only on the second principle of thermodynamics, on the other hand, the entropy functional formalism \cite{Padmanabhan1, Padmanabhan2} is able to reproduce, not only general relativity (see also \cite{Hammad1}) and Einstein-Cartan theory \cite{Hammad2}, as well as the main modified gravity theories \cite{Padmanbhan3, Hammad3}, but also cosmic expansion \cite{Hammad4} and black hole thermodynamics, i.e. the Bekenstein-Hawking entropy formula \cite{Padmanabhan1, Hammad1}. It would hence be of significant interest to investigate also within this formalism the possibility of the appearance of this famous logarithmic correction and how, if any, would the appearance of this term affect the resulting gravitational theory based on the formalism.

As we shall see in this work, the logarithmic correction to the Bekenstein-Hawking formula arises in a natural way for stable black holes within the entropy functional formalism thanks to the correction of the standard micro-canonical entropy around the equilibrium introduced recently by S. Das \textit{et al.} \cite{Das2} (see also \cite{Faizal,Pourhassan}). Since the entropy functional formalism is based on the familiar concept of entropy used in the study of elastic solids (see e.g., \cite{Landau}) and the second principle of thermodynamics, which states that this entropy should always be maximal, the universal entropy correction found in Ref.~\cite{Das2} could straightforwardly be incorporated into the formalism. Indeed, as we show here, associating this corrected expression of entropy with the deformation vector field of a spacetime region yields the famous logarithmic correction for stable black holes' entropy. More important, however, is the fact that the gravitational theory that comes out of the formalism thus modified remains, at each order in the spacetime curvature, unaltered. But, as we shall see, the additional terms that come with the corrected entropy allows one to impose non-trivial constraints on the modified gravity theory resulting from the formalism. Subtleties related to the fact that the approach originally introduced in Ref.~{\cite{Das2}} is only valid for a thermodynamical system at equilibrium or, equivalently, for a black hole with positive heat capacity, is also dealt with.

The outline of this paper is as follows. In Sec.~\ref{sec:2}, we give a brief reminder of the philosophy behind the entropy functional formalism as well as a brief description of the corrected micro-canonical entropy introduced in Ref.~\cite{Das2} that we shall adopt here. In Sec.~\ref{sec:3}, we incorporate this new entropy expression into the entropy functional formalism and show that this modification does not alter in anyway the spacetime dynamics one obtains by imposing the second principle of thermodynamics on this new form of the functional, and that when applied to a spherically symmetric black hole, satisfying the condition of stability, the corresponding entropy acquires the famous corrected logarithmic term. In Sec.~\ref{sec:4}, we examine the case of an $\mathcal{F}(R)$-gravity and deduce, using the second principle of thermodynamics, constraints on this class of modified gravity theories. We end this paper with a brief conclusion.

\section{The Entropy Functional Formalism and the Modified Entropy Formula}\label{sec:2}
In the entropy functional formalism one views spacetime as a continuum medium subject to deformations, just as an elastic solid is subject to elastic deformations. The deformations of spacetime in this formalism are defined, in analogy with the standard elasticity theory \cite{Landau}, via the deformation vector field $u^{\mu}$ which quantifies the difference between two configurations of spacetime, $\overline{x}^{\mu}$ after deformation and $x^{\mu}$ before deformation: $u^{\mu}=\overline{x}^{\mu}-x^{\mu}$. Then, still in analogy with standard elasticity theory, an entropy functional is associated with these deformations. The functional is quadratic in the field $u^{\mu}$ as well as in its first derivatives $\nabla u^{\mu}$. Also, as justified in Refs.~\cite{Hammad2, Hammad3}, in addition to these quadratic terms the functional could contain hybrid terms of the form $u\nabla u$. The entropy functional is expressed in the following general integral form
\begin{equation}\label{1}
{\cal S}=\int_{\mathcal{M}}\mathrm{d}^{4}x\sqrt{-g}\;\mathfrak{s}(u,\nabla u)
\end{equation}
where $\sqrt{-g}\;\mathfrak{s}(u^{\mu},\nabla u^{\mu})$ is the entropy functional density, integrated over that spacetime region of interest $\mathcal{M}$. A variational principle is then applied to this entropy functional in accordance with the second principle of thermodynamics which states that the entropy of any isolated system at equilibrium should be maximal. However, whereas the usual action principle for gravity relies on a variation of the action with respect to the metric of spacetime, the variational principle used in the entropy functional formalism relies on a variation of the integral with respect to the deformation vector field $u^{\mu}$. Indeed, the philosophy behind the whole approach is that the dynamics of spacetime, that is the dynamical equation that governs the spacetime metric, represents precisely that configuration of spacetime that renders maximum the entropy associated with the deformations $u^{\mu}$.

The spacetime dynamics one obtains depends actually on what structure one has inside the density $\sqrt{-g}\;\mathfrak{s}(u^{\mu},\nabla u^{\mu})$. The more curvature terms one includes in the construction of this density, the higher-curvature corrections with respect to general relativity appear in the final dynamical equations of the metric. The most general construction of the entropy functional density in which, not only higher powers of the curvature tensors might be present, but also a scalar field $\phi$ is allowed in, is the following \cite{Hammad3}:
\begin{align}\label{2}
\mathfrak{s}(u,\nabla u)&=\Phi_{\mu\nu\rho\sigma}\nabla^{\mu}u^{\nu}\nabla^{\rho}u^{\sigma}+\Psi_{\mu\nu\rho}u^{\mu}\nabla^{\nu}u^{\rho}
+\Pi_{\mu\nu}u^{\mu}u^{\nu},
\end{align}
where $\Phi_{\mu\nu\rho\sigma}$, $\Psi_{\mu\nu\rho}$, and $\Pi_{\mu\nu}$ are fourth-, third-, and second-order tensors, respectively. These tensors are built from the Riemann tensor, its contractions and derivatives, and/or a scalar field $\phi$ and its derivatives. The variation of (\ref{1}) with respect to the field $u^{\mu}$, using the general from (\ref{2}) of the functional density, gives the corresponding scalar-tensor theory of gravity which is compatible with that order in the curvature chosen inside the three tensors above.

Now, it is clear that the whole approach relies on the functional integral (\ref{1}) which, in turn, is borrowed from the usual elasticity theory of three-dimensional elastic solids. Therefore, integral (\ref{1}) is actually equivalent simply to the statement, $\mathcal{S}=S_{0}$, which just identifies the entropy functional at equilibrium with the micro-canonical entropy $S_{0}$ associated with the deformations $u^{\mu}$ at the equilibrium temperature $1/\beta_{0}$ \footnote{We shall, throughout the paper, set Boltzmann's constant equal to unity. Therefore, $\beta=1/T$}. However, it has recently been shown in Ref.~\cite{Das2} that a more correct expression for the micro-canonical entropy $\mathcal{S}$ of any thermodynamical system fluctuating around the equilibrium should instead be written in the form
\begin{equation}\label{3}
\mathcal{S}=\ln\rho(E)=S_{0}-\frac{1}{2}\ln\left(\frac{\partial^{2}S(\beta)}{\partial\beta^{2}}\right)_{\beta=\beta_{0}}+...,
\end{equation}
where $\rho(E)$ is the density of states corresponding to the energy $E$ and the ellipsis stands for higher order terms. The appearance of the additional logarithmic term in the latter expression was explained in Ref.~\cite{Das2} as due to fluctuations of the system about the equilibrium whose exact entropy at any temperature $(1/\beta)$ is $S(\beta)$. Indeed, the density of states $\rho(E)$ appears in the partition function $Z(\beta)=\int\rho(E)\exp(-\beta E)\mathrm{d}E$ of the canonical ensemble describing the system, whereas the partition function itself appears inside the exact since $S(\beta)=\ln Z(\beta)+\beta E$ for any temperature $1/\beta$ of the system. Then, by simply Taylor-expanding $S(\beta)$ up to the second order in $\beta$ as, $S(\beta)=S(\beta_{0})+\frac{1}{2}(\beta-\beta_{0})^{2}(\partial^{2}S/\partial\beta^{2})_{\beta=\beta_{0}}$, one has access to entropy fluctuations around the equilibrium $S_{0}:=S(\beta_{0})$ reached at temperature $1/\beta_{0}$. Thanks to the previous relation between $S(\beta)$ and $Z(\beta)$, these fluctuations get transmitted inside the partition function $Z(\beta)$ and then, after inverting the above integral relating $Z(\beta)$ to the density of states $\rho(E)$, the latter also gets infected by these fluctuations. The final result is that the micro-canonical entropy $\mathcal{S}=\ln\rho(E)$ acquires, as a leading correction, the logarithmic term given in \ref{3}. For the detailed analytic steps of the derivation of this result, see Ref.~\cite{Das2}.

\section{Gravity and Stable Black Holes Thermodynamics from the Corrected Entropy Formula}\label{sec:3}
Since in the entropy functional formalism it is implicit that the entropy associated with the deformations $u^{\mu}$ of spacetime should always be maximal, this is equivalent to taking spacetime as an isolated system at equilibrium. Therefore, the above formula (\ref{3}), supposed to be valid for any thermodynamical system fluctuation around the equilibrium, should also be adopted as the entropy to be associated with the deformations $u^{\mu}$ of spacetime viewed as a continuum thermodynamical system. Hence, based on this argument the more correct form we should investigate within the entropy functional formalism that would take into account thermal fluctuations of the system around equilibrium should also include this logarithmic correction. That is, the entropy functional associated with the field $u^{\mu}$ should be of the form
\begin{equation}\label{4}
\mathcal{S}=\int_{\mathcal{M}}\mathrm{d}^{4}x\sqrt{-g}\left[\mathfrak{s}(u,\nabla u)\right]_{\beta_{0}}-\frac{1}{2}\ln\int_{\mathcal{M}}\mathrm{d}^{4}x\sqrt{-g}\left[\frac{\partial^{2}\mathfrak{s}(u,\nabla u)}{\partial\beta^{2}}\right]_{\beta_{0}}.
\end{equation}
Now, as usual, we should impose the variational principle on this functional to guaranty that the resulting entropy is indeed extremum for every deformation vector field $u^{\mu}$, and find, in the resulting equations obeyed by the spacetime metric, that configuration of spacetime that satisfies this requirement. The variation of the above integral reads
\begin{equation}\label{5}
\delta\mathcal{S}=\int_{\mathcal{M}}\mathrm{d}^{4}x\sqrt{-g}\left[\left(1-\frac{1}{2I}\frac{\partial^{2}}{\partial\beta^{2}}\right)
\;\delta\mathfrak{s}(u,\nabla u)\right]_{\beta_{0}},
\end{equation}
where $I=\int_{\mathcal{M}}\mathrm{d}^{4}x\sqrt{-g}(\partial^{2}\mathfrak{s}/\partial\beta^{2})_{\beta_{0}}$. It is clear now that the condition for the vanishing of this latter expression, for any vector field $u^{\mu}$ and any temperature $1/\beta_{0}$, reduces to the vanishing of the second factor inside the square brackets of the integrand, i.e. $\delta\mathfrak{s}=0$. Therefore, the variation of (\ref{4}) would give the same constraints on spacetime, i.e. the metric equations of motion, we would have obtained if we had discarded the second logarithmic term there. However, when inserting the resulting equations of motion for the metric back into (\ref{4}), both bulk integrals there become boundary integrals and the logarithmic term subsists. Indeed, integration by parts of (\ref{4}) gives
\begin{align}\label{6}
\mathcal{S}&=\int_{\mathcal{\partial M}}\mathrm{d}^{3}x\sqrt{|h|}\left(n^{\mu}\Phi_{\mu\nu\rho\sigma}u^{\nu}\nabla^{\rho}u^{\sigma}
+\frac{1}{2}n^{\mu}\Psi_{\nu\mu\rho}u^{\nu}u^{\rho}\right)_{\beta_{0}}\nonumber
\\&\quad-\frac{1}{2}\ln\int_{\mathcal{\partial M}}\mathrm{d}^{3}x\sqrt{|h|}\left[\frac{\partial^{2}}{\partial\beta^{2}}n^{\mu}\left(\Phi_{\mu\nu\rho\sigma}u^{\nu}\nabla^{\rho}u^{\sigma}
+\frac{1}{2}\Psi_{\nu\mu\rho}u^{\nu}u^{\rho}\right)\right]_{\beta_{0}},
\end{align}
where the general expression (\ref{2}) of $\mathfrak{s}$ has been used. $h$ is the three-metric induced on the boundary $\partial\mathcal{M}$, and $n^{\mu}$ is the normal vector on the latter.

To apply this 'on-shell' entropy to the calculation of the entropy of black holes, the prescription is to follow the three following steps \cite{Padmanabhan1}. First, go to the near-horizon Rindler geometry and separate in (\ref{6}) the time from the spatial integral, then choose a periodic time that lies within the interval $[0,\beta]$:
\begin{equation}\label{7}
\int_{\mathcal{H}}\mathrm{d}^{3}x\sqrt{|h|}\rightarrow\int_{0}^{\beta}\mathrm{d}t\int_{L_{\perp}}\mathrm{d}\Sigma,
\end{equation}
where $\mathrm{d}\Sigma$ is the surface element on the space $L_{\perp}$ orthogonal to the Rindler near-horizon geometry extracted from $\mathrm{d}s^{2}=-\kappa^{2}N\mathrm{d}t^{2}+\mathrm{d}N^{2}+\mathrm{d}l_{\perp}^{2}$. Here, $\kappa$ is the surface gravity of the horizon located at $N=0$, and $\mathrm{d}l_{\perp}^{2}$ is the line element in the space $L_{\perp}$. Next, assume that the deformation field $u^{\mu}$ is always normal to a space-like hyper-surface, in accordance with the many fingered time slicing, and satisfies $n^{\mu}u_{\mu}=0$ on the null-surface of the event horizon. Finally, take into account the fact that $u^{\nu}\nabla_{\nu}u^{\mu}=a^{\mu}$ is the four-acceleration associated with the deformation vector field $u^{\mu}$, satisfying $Nn^{\mu}a_{\mu}\rightarrow\kappa$ on the event horizon. The periodicity $\beta$ of $t$ is a requirement to avoid the conical singularity that would otherwise form in the $(t,N)$-plane of the Rindler near-horizon space. Indeed, the latter would exhibit a deficit in closure in the two-plane spanned by $t$ and $N$ unless $t$ is taken periodic with period $2\pi\kappa^{-1}$. On the other hand, it is well known that the two-point function $\langle\phi_{2}|\exp(iH(t_{2}-t_{1}))|\phi_{1}\rangle$ of quantum field theory becomes the statistical partition function $Z=\mathrm{Tr}\exp(-\beta H)$ after a Wick rotation $t_{2}-t_{1}\rightarrow i\tau$ and identification of $\tau$ with $\beta$. Therefore, these two arguments taken together show that for consistency one should actually take $Nn^{\mu}a_{\mu}\rightarrow 2\pi\beta^{-1}$ at the event horizon.

As it is, integral (\ref{6}) does not allow one to find a specific value for entropy of a black hole. It is only after one chooses the tensors $\Phi_{\mu\nu\rho\sigma}$ and $\Psi_{\mu\nu\rho}$ at a given order in the curvature, or equivalently, at a given power of the gravitational constant $G$ \cite{Hammad3}, that one could perform all the details outlined in the three steps indicated above. For the sake of concreteness, let us choose here first the simplest structure for these tensors; namely, let us limit ourselves to the lowest order in the curvature, a case which gives general relativity. The latter is recovered indeed for $\Phi_{\mu\nu\rho\sigma}=(g_{\mu\sigma}g_{\nu\rho}-g_{\mu\nu}g_{\rho\sigma})/8\pi G$ and $\Psi_{\mu\nu\rho}=0$. For this case, and after performing the three steps outlined in the previous paragraph, integral (\ref{6}) reduces to
\begin{equation}\label{8}
\mathcal{S}=\int_{0}^{\beta_{0}}\mathrm{d}t\frac{A}{4\beta_{0}G}
-\frac{1}{2}\ln\int_{0}^{\beta_{0}}\mathrm{d}t\left[\frac{\partial^{2}}{\partial\beta^{2}}\left(\frac{A}{4\beta G}\right)\right]_{\beta=\beta_{0}}
=\frac{A}{4G}-\frac{1}{2}\ln\frac{A}{2\beta_{0}^{2}G},
\end{equation}
where $A$ is the area of the event horizon. Thus we see that the modification (\ref{4}) of the entropy functional formalism allows one to recover the famous logarithmic correction to black holes' entropy, provided of course that these black holes have a positive heat capacity, i.e. stable, for the approach to be viable. The consistency of this last result can be verified by noticing that since $\beta_{0}^{2}$ is proportional to $S_{0}$, as it follows from the zeroth-order of the unperturbed system \footnote{Here we use the fact that once the Bekenstein-Hawking entropy $S_{0}$ is found in terms of the black hole's mass $M$, one uses $T\mathrm{d}S=\mathrm{d}M$ to deduce that the Hawking temperature $T_{0}$ is such that $S_{0}\propto T^{-2}_{0}=\beta_{0}^{2}$ \cite{Hammad1}.}: $\mathcal{S}=S_{0}=A/4G\propto\beta_{0}^{2}$, the last term in (\ref{8}) vanishes and we are left, up to an unimportant constant, with $\mathcal{S}=S_{0}$.

\section{Constraining the Modified Gravity}\label{sec:4}
The utility of the entropy functional formalism resides in the fact that the more powers of Newton's constant $G$ one allows inside the general expression (\ref{2}), the higher curvature terms one recovers at the end inside the resulting modified gravity theory \cite{Hammad3}. Therefore, since, as we saw in the previous section, the formalism yields the same modified gravity theory even when used in its modified version (\ref{4}), the natural question that arises is whether the resulting modified gravity would be effected in any way and to what extent.

The answer to the above question can be extracted from identity (\ref{6}) since the latter appeals to the full structure of the entropy functional. In order to use (\ref{6}), however, we should first decide upon the order of approximation we wish to achieve in the final modified gravity theory. Given that the determination of the three tensors $\Phi_{\mu\nu\rho\sigma}$, $\Psi_{\mu\nu\rho}$, and $\Pi_{\mu\nu}$ from the formalism becomes tedious when going up in curvature terms, and given that the first-order in curvature was already treated in Ref.~\cite{Hammad3}, we shall restrain ourselves here to that order. Upon imposing the extremum condition on (\ref{4}) one finds the structure of the three tensors from which one deduces the modified gravity's field equations. The field equations obtained in Ref.~\cite{Hammad3} at this order were the equations that would result from the gravitational Lagrangian $\mathcal{F}(R, R_{\mu\nu}R^{\mu\nu}, R_{\mu\nu\rho\sigma}R^{\mu\nu\rho\sigma}, \phi)=a\mathcal{R}^{2}+\mathcal{M}(\phi)R_{\mu\nu\rho\sigma}R^{\mu\nu\rho\sigma}+\mathcal{D}(\phi,R)+\mathcal{U}(\phi)$, where $a$ is an undetermined constant, $\mathcal{R}^{2}$ is the Gauss-Bonnet invariant $R^{2}-4R_{\mu\nu}R^{\mu\nu}+R_{\mu\nu\rho\sigma}R^{\mu\nu\rho\sigma}$, whereas $\mathcal{M}(\phi)$ and $\mathcal{U}(\phi)$ are functionals of the scalar field $\phi$ alone in contrast to $\mathcal{D}(\phi,R)$ which depends also on the Ricci scalar $R$.

For a black hole solution of the form $\mathrm{d}s^{2}=-f(r)\mathrm{d}t^{2}+\mathrm{d}r^{2}/f(r)+r^{2}(\mathrm{d}\theta^{2}+\sin^{2}\theta\mathrm{d}\phi^{2})$ within the above modified gravity $\mathcal{F}(R, R_{\mu\nu}R^{\mu\nu}, R_{\mu\nu\rho\sigma}R^{\mu\nu\rho\sigma}, \phi)$, with $f(r)$ a smooth function such that $f(r_{\mathcal{H}})=0$, the 'on-shell' entropy (\ref{6}) takes on, after applying again the three steps outlined in the previous section (see \cite{Hammad3}), the following expression:
\begin{equation}\label{9}
\mathcal{S}=\left(\int_{0}^{\beta_{0}}-\frac{1}{2}\ln\int_{0}^{\beta_{0}}\frac{\partial^{2}}{\partial\beta^{2}}\right)
\left[\frac{A}{4\beta G}\left(\frac{a}{r_{\mathcal{H}}^{2}}+\frac{\partial\mathcal{D}}{\partial R}-2\mathcal{M}\partial^{2}_{r}f\big|_{\mathcal{H}}\right)\right]_{\beta=\beta_{0}}\mathrm{d}t,
\end{equation}
where the square brackets should be evaluated at $\beta=\beta_{0}$ after performing the second partial derivative with respect to $\beta$ from the left. Finally, the evaluation of the two integrals permits to find the following black hole entropy:
\begin{equation}\label{10}
\mathcal{S}=\frac{A}{4G}\left(\frac{a}{r_{\mathcal{H}}^{2}}+\frac{\partial\mathcal{D}}{\partial R}-2\mathcal{M}\partial^{2}_{r}f\big|_{\mathcal{H}}\right)
-\frac{1}{2}\ln\frac{A}{2\beta_{0}^{2}G}-\frac{1}{2}\ln\left(\frac{a}{r_{\mathcal{H}}^{2}}+\frac{\partial\mathcal{D}}{\partial R}-2\mathcal{M}\partial^{2}_{r}f\big|_{\mathcal{H}}\right).
\end{equation}

The first thing we satisfactorily notice here is that the logarithmic corrections to entropy in the modified gravity theory obtained are due to the gravitational Lagrangian and not only to the area of the black hole's event horizon as was the case in (\ref{8}) found within general relativity. In other words, the logarithmic corrections can produce the modified theories of gravity in the entropy functional formalism. The result (\ref{8}) can be recovered from (\ref{10}) by choosing $a=0$, $\mathcal{M}(\phi)=0$, $\mathcal{D}(\phi,R)=R$, and $\mathcal{U}(\phi)=0$, in which case one recovers the Hilbert Lagrangian. The content of the parentheses in (\ref{10}) then reduces to $1$ and the the second logarithm vanishes.

The second thing we notice from (\ref{10}) is that the sign of the contributions to entropy from modified gravity via the logarithmic corrections changes, depending on the gravitational theory one has and on the value of the Ricci scalar of the Universe. But, since entropy should not be negative and should not be decreasing, in accordance with the second principle of thermodynamics, we should impose, besides the condition $a/r_{\mathcal{H}}^{2}+\partial\mathcal{D}/\partial R>2\mathcal{M}\partial^{2}_{r}f\big|_{\mathcal{H}}$, another condition that guaranties that $\partial\mathcal{S}/\partial R\leq0$ on the event horizon $\mathcal{H}$. This last condition is motivated by the fact that as the Universe expands, the Ricci scalar decreases, but entropy should increase.

However, in order to use (\ref{10}) to translate these conditions into detailed inequalities, we would need to have the explicit expressions of the radius $r_{\mathcal{H}}$ as well as the function $f(r)$ for the black hole solution, which are both unknown within the $\mathcal{F}(R,R_{\mu\nu}R^{\mu\nu},R_{\mu\nu\rho\sigma}R^{\mu\nu\rho\sigma},\phi)$ theory above. Therefore, let us limit our investigation here to the simple case of an $\mathcal{F}(R)$-gravity theory, which corresponds here to the case $a=0=\mathcal{M}$ and $\mathcal{D}(R)\equiv\mathcal{F}(R)$. The field equations as well as the black hole solution of $\mathcal{F}(R)$-gravity are already well established \cite{Briscese,Cruz1,Cruz2}. First, the Ricci scalar $R$ of an $\mathcal{F}(R)$-gravity theory should be constant (unless $\mathcal{F}(R)\propto R^{2}$). Let us denote the constant Ricci scalar by $R_{0}$. Second, a black hole solution in an $\mathcal{F}(R)$-gravity theory is again of the form $\mathrm{d}s^{2}=-f(r)\mathrm{d}t^{2}+\mathrm{d}r^{2}/f(r)+r^{2}(\mathrm{d}\theta^{2}+\sin^{2}\theta\mathrm{d}\phi^{2})$ with the radius $r_{\mathcal{H}}$ of the black hole given by $f(r_{\mathcal{H}})=1-2GM/r_{\mathcal{H}}-R_{0}r_{\mathcal{H}}^{2}/12=0$. Therefore, we have $A=4\pi r_{\mathcal{H}}^{2}$ and, since the constant $R_{0}$ was arbitrary, we have for all $R$
\begin{equation}\label{11}
\frac{\partial A}{\partial R}=\frac{\partial A}{\partial r_{\mathcal{H}}}\frac{\partial r_{\mathcal{H}}}{\partial R}=\frac{A^{2}}{24\pi-3RA/2}.
\end{equation}

Before we proceed further, we would like to make an important remark here. As we have emphasized it above, the logarithmic corrections introduced in Ref.~\cite{Das2} are only valid for systems at equilibrium. However, the black hole solution we are considering here is a black hole in an expanding Universe, i.e. a Schwarzschild-de Sitter black hole. It is, however, known that this type of black hole evaporates and, therefore, has a negative heat capacity. What we need though, in order to be able to apply the logarithmic corrections, is a black hole in a de Sitter spacetime which still has a positive heat capacity. Fortunately, such a possibility has been pointed out by Bousso and Hawking in Ref.~\cite{Bousso}. The authors showed there that the Nariai black hole \cite{Nariai}, which is a degenerate Schwarzschild-de Sitter black hole at the limit where the black hole's horizon and the cosmological horizon coincide, might anti-evaporate and therefore possess a positive heat capacity. On the other hand, given that the black hole solution we are assuming is a solution in an $\mathcal{F}(R)$-modified gravity, we also need an $\mathcal{F}(R)$-gravity black hole solution that has a positive heat capacity in a de Sitter spacetime. Fortunately here also, such a possibility exists and have been studied in Ref.~\cite{Nojiri,Sebastiani}. Consequently, the black hole solution we assumed above should be a Nariai black hole for the logarithmic corrections to be legitimate. This restriction, however, will not put any limitation on the generality of the conclusions we will reach below, for the equations used are supposed to be valid for every curvature scalar $R$; it is just that these are derived using a well chosen system so that the logarithmic corrections approach can be applied.

Now, since Eq.~(\ref{10}) is valid at the event horizon for whatever value the Ricci scalar might possess, we will partially differentiate Eq.~(\ref{10}) with respect to an arbitrary $R$ and impose $\partial\mathcal{S}/\partial R\leq0$. We find
\begin{equation}\label{12}
\frac{\partial\mathcal{S}}{\partial R}=\left(\frac{A}{4G}-\frac{1}{2\mathcal{F}_{,R}}\right)\left(\frac{\partial A}{A\partial R}\mathcal{F}_{,R}+\mathcal{F}_{,RR}\right)\leq0,
\end{equation}
where $\mathcal{\mathcal{F}}_{,R}$ and $\mathcal{F}_{,RR}$ stand for the first and second derivatives of $\mathcal{F}$ with respect to $R$. Since from (\ref{10}) we already know that $\mathcal{F}_{,R}>0$, and from \cite{Nojiri2,Felice} that $\mathcal{F}_{,RR}>0$, we deduce that one possibility for inequality (\ref{12}) to be satisfied is to have the first parenthesis negative provided that in the second, the factor of $\mathcal{F}_{R}$ is positive. From (\ref{11}) we have $\partial A/\partial R>0$ for every event horizon $A$ of a Nariai black hole provided that $RA<16\pi$, that is, $r_{\mathcal{H}}^{2}<4/R$. To translate this in terms of the mass $M$ of the black hole, just use the equation $f(r_{\mathcal{H}})=0$ to find that the mass $M$ of the Nariai black hole and the degenerate horizon at $r_{\mathcal{H}}$ should satisfy $GM<r_{\mathcal{H}}/3$ to be guaranteed to have $\partial A/\partial R>0$. The condition (\ref{12}) would then simply be equivalent in this case to $0<\mathcal{F}_{,R}\leq2G/A$ and $\mathcal{F}_{,RR}>0$. This cannot be satisfied for a Nariai black hole, however, because the two positive roots of the equation $f(r_{\mathcal{H}})=0$ meet each other at $r_{\mathcal{H}}$ such that $(\partial_{r}f)(r_{\mathcal{H}})=0$. This gives thereby the condition $r_{\mathcal{H}}^{2}=4/R$. Hence, this first possibility cannot not be satisfied and equation (\ref{11}) cannot not even be defined in this case.

The second possibility for (\ref{12}) to be satisfied is to have the first parenthesis positive whereas the second parenthesis negative. We actually find for the extremal value of $A=4\pi r_{\mathcal{H}}^{2}=16\pi/R$, where the two horizons coincide, that $\partial A/\partial R<0$. Therefore, the second parenthesis in inequality (\ref{12}) could indeed be negative while the first is positive. By substituting this extremal value of $A$ in (\ref{12}), this second possibility then translates into the following two conditions
\begin{equation}\label{13}
\mathcal{F}_{,R}\geq\frac{GR}{8\pi}\qquad\mathrm{and}\qquad\frac{R\mathcal{F}_{,RR}}{\mathcal{F_{,R}}}<1.
\end{equation}
Thus we see that thermodynamics is also capable of imposing nontrivial constraints on the modified gravity theory at hand. The first condition is more restrictive than the simple more familiar condition $\mathcal{F}_{,R}>0$ found in the literature. The second condition is also more restrictive than the simple one $\mathcal{F}_{,RR}>0$, and both have remarkably already been obtained from a completely different investigation; more specifically, they have been obtained in the literature from cosmological requirements, see e.g. \cite{Felice}.

\section{Conclusion}
We have used the entropy functional formalism, for its capacity to reproduce general relativity as well as well-known theories of modified gravity, and incorporated into it the recently universal first-order entropy correction proposed to apply for all thermodynamical systems fluctuating around the equilibrium by S. Das \textit{et al} \cite{Das2}. We found that this combination did not alter the gravitational theories one would obtain from the formalism before including this corrected entropy, whereas the result for such black hole thermodynamics is modified. The modification to black hole thermodynamics found here is in agreement in its form, i.e. displaying terms logarithmic in area, with what is found elsewhere using other approaches to quantum gravity.

The fact that the entropy functional formalism gives rise to modified gravity and simultaneously to black holes thermodynamics turned out to be a bonus that helps deduce constraints on the modified gravity. Since only in $\mathcal{F}(R)$-gravity do we have a precise black hole solution, we have restrained ourselves to this case and deduced constraints to be imposed on the theory. We found that these constraints are nothing but the
constraints imposed on $\mathcal{F}(R)$-gravity on cosmological grounds. It is remarkable that the same constraints also arise from thermodynamics.

We would like to stress here the fact that the constraints we found are obtained by applying the logarithmic corrections approach to a very specific system, namely, the Nariai black hole, for the approach is only applicable on stable black holes as already emphasized by the authors in Ref.~\cite{Das2}. This restriction does not, however, entail any limitation on the generality of the inequalities (\ref{13}). Indeed, the fact that a particular system, for which the method is viable, must be chosen among all the systems present within the expanding Universe simply reflects the limitations of the method but in no way does it restrict the validity of the results deduced from the application of the method to systems for which it remains viable. Furthermore, the fact that both the $\mathcal{F}(R)$-gravity model as well as the Ricci scalar were not chosen but left arbitrary inside the equations, might only point to the universality of the results.

Now it is satisfactory to have been able to recover a result on black hole thermodynamics on which different axis of research converge and yet using the same formalism that served to reproduce classical general relativity and which is capable of giving rise to modified gravity. It is however intriguing at the same time that the black hole result is recovered without the machinery of the more sophisticated recent quantum approaches such as the generalized uncertainty principle which gives also logarithmic corrections (see e.g., \cite{Awad, Medved, Gangopadhyay, Tawfik1, Tawfik2}). Using Wick rotation and periodic times in our formalism is of course related to quantum theory, but the logarithmic term itself does not have any obvious origin in quantum theory. Indeed, the logarithmic correction we adopted here from the work of S. Das \textit{et al.} \cite{Das2} is supposed to be valid for any fluctuation around the equilibrium of a thermodynamic system and not just quantum systems. Having applied it successfully to the entropy functional formalism and derived with it a result inherently quantum mechanical suggests that it is likely that the entropy functional formalism might have real roots in the quantum theory of gravity, and hence, real chances to play a significant role in the search for the latter.
\\

\end{document}